\documentclass[12pt,preprint]{aastex}

\newcommand{\be}{\begin{equation}}
\newcommand{\ee}{\end{equation}}
\newcommand{\bea}{\begin{eqnarray}}
\newcommand{\eea}{\end{eqnarray}}
\newcommand{\bean}{\begin{eqnarray*}}
\newcommand{\eean}{\end{eqnarray*}}

\shorttitle{Making Hot Jupiters}
\shortauthors{Shara, Hurley \& Mardling}

\slugcomment{Submitted to the Astrophysical Journal}

\begin{document}

\title{Dynamical Interactions Make Hot Jupiters in Open Star Clusters}

\author{Michael M. Shara}
\affil{Department of Astrophysics, 
       American Museum of Natural History, \\
       Central Park West at 79th Street, 
       New York, NY 10024}
\email{mshara@amnh.org}

\author{Jarrod R. Hurley}
\affil{Centre for Astrophysics \& Supercomputing, PO Box 218, \\ 
        Swinburne University of Technology, Victoria 3122, Australia}
\email{jhurley@swin.edu.au}

\author{Rosemary A. Mardling}
\affil{Department of Mathematics, PO Box 28M, \\ 
        Monash University, Victoria 3800, Australia}
\email{Rosemary.Mardling@sci.monash.edu.au}

\begin{abstract}

Explaining the origin and evolution of exoplanetary ``hot Jupiters"  remains a significant challenge. One possible
mechanism for their production is planet-planet interactions, which produces hot Jupiters from planets born far from their host stars but near their dynamical stability limits. In the much more likely case of planets born far from their dynamical stability limits, can hot Jupiters can be formed in star clusters? Our N-body simulations answer this question in the affirmative, and show that hot Jupiter formation is not a rare event, occurring in $\sim1\% $ of star cluster planetary systems. We detail three case studies of the dynamics-induced births of hot Jupiters on highly eccentric orbits that can only occur inside star clusters. The hot Jupiters' orbits bear remarkable similarities to those of some of the most extreme exoplanets known: HAT-P-32 b, HAT-P-2 b, HD 80606 b and GJ 876 d. If stellar perturbations formed these hot Jupiters then our simulations predict that these very hot, inner planets are often accompanied by much more distant gas giants in highly eccentric orbits.

\end{abstract}

\keywords{stellar dynamics---methods: N-body simulations---
          planetary systems---open clusters and associations: general}

\section{Introduction}
\label{s:intro}

The detection of exoplanets has revolutionized planetary astronomy \citep{wol92,may95,bat13}.
The highly eccentric and/or compact orbits of many of the newly discovered exoplanets \citep{mar03}
are no less extraordinary than the discoveries of the exoplanets themselves.
Explaining the remarkable orbits of exoplanets remains one of
the most important problems in this emerging field. Even if ``hot Jupiters" and other exotic exoplanets are 
but a small fraction of all the planetary systems in our Galaxy,
accounting for them is a key part of any successful theory of planetary formation and evolution. 

Most attempts at explaining the observed distribution of exoplanets' orbits
have invoked dynamical effects intrinsic to newly forming 
solar systems. Interactions between nascent planetesimals, 
planets and/or the disks and gas out of which they form
can all cause migration of planetary orbits \citep{gol80,lin96,ras96,mar03} and 
misaligned hot Jupiters \citep{thi11}, with observational evidence recently presented by \citet{dej12}. 
Diversity may arise in strong star-planet interactions \citep{lin08}.
Binary stellar companions can also change the
orbits of planets \citep{egg03}. Perhaps the most challenging observation 
to explain is the observed plot of eccentricities, $e$, versus semi-major axes, $a$, 
of exoplanets \citep{mar03}. No single mechanism has yet been shown 
to reproduce this distribution. Multiple planet scattering may 
be promising \citep{for03,for06}, but it fails to produce the frequency 
of giant planets with semi-major axes smaller than about $1\,$ Astronomical Unit (AU).  \citet{ada05} has demonstrated 
that combining dynamical scattering with tidal interactions from 
a circumstellar disk can populate the observed range of $a$ and $e$ with planets. 
Is this degree of complexity required to explain exoplanetary orbits? 
Does this suggest that our own solar system's nearly circular orbits are 
rather unusual? Before answering yes, it is important to investigate other orbit-altering mechanisms 
(recently reviewed by \citet{hao13}). 

Many (and perhaps most) stars are born in or near star clusters or associations, 
eventually escaping as their host clusters dissolve \citep{lad03,kru12}. 
Many clusters either do not escape their birth environments or dissolve within 20-50 Myr of formation \cite{lad03,god10}, 
but the existence of the nearby Pleiades (at roughly 100 Myr), and the Hyades and Presepe clusters (about 600 Myr each) 
demonstrates that some long-lived clusters survive. Planets have been detected in star clusters \citep{lov07,sat07}. 
The detection of two planets smaller than Neptune, in the billion year old open cluster NGC 6811 \citep{mei13},
suggest that planets are as common in open star clusters as they are in the field.

\citet{qui12} have reported the discovery of two hot Jupiters in Praesepe. The discovery of three planetary companions to M67 stars 
includes two hot Jupiters \citep{bru14}. \citet{qui14} have reported an eccentric hot Jupiter in the Hyades open cluster.
It is natural to ask if interactions with passing stars might be responsible for significantly
changing the orbits of some planets in clusters, and perhaps producing hot Jupiters. The importance of passing stars in perturbing planetary bodies should increase with
stellar density, the cross sections of planetary systems and the lifetimes of planetary systems inside star clusters.
\citet{lau98,spu03,ada01,ada06} and \citet{ada15} have shown via scattering experiments that the orbits of planetary systems in star clusters 
must be changed during encounters with passing stars.

\citet{fue97,hur02,spu09} and \citet{par12} have shown, with full $N$-body star cluster plus 
planet simulations, that the orbits of {\it single} planets in clusters are dramatically affected by stellar encounters. 
The cumulative effects of passing stars change the orbits of single gas giants in star clusters, 
liberating some planets while leaving others bound to their host stars for many Myr. 
The evolution of multi-planet systems in star clusters is a much more difficult computational task.
It is important to simulate multi-planet systems both because they exist and because planet-planet interactions may be 
as important as stellar encounters in modifying orbits \citep{hsm08,mal11,bol12}. A significant advance was recently made by \cite{hao13} and by \citet{ada15},
who used Monte Carlo scattering experiments to conclusively demonstrate that flybys strongly affect the evolution of multi-planet systems in 
dense open clusters. 
 
The goal of this paper is to check if the hottest observed Jupiters can be formed in multi-planet systems in a star cluster.
We report here that it is remarkable but true that we can reproduce the orbits of the hottest known exoplanets
with ease in a system with just two gas giants.  Our simulation method and initial conditions are detailed in Section 2. The development
of three extreme hot Jupiters is described in Section 3. These planets' orbits are compared to 
the orbits of some of the most extreme hot Jupiters known in Section 4. 
We briefly summarize our results in Section 5.

\section{Simulation Method}
\label{s:method}

Self-consistently modeling the evolution of solar systems in star clusters
is computationally challenging. The orbits of planets must be followed 
for many Myr at the same time as those of the thousands of stars in the cluster.
To model the dynamical evolution of star clusters as well as the single and binary stars
in the cluster we use the Aarseth {\tt NBODY6} code \citep{aar99,hur01}. Tidal circularisation is included in the 
modelling via our usual method for treating binaries \citep{hur01}. 
The simulations reported here were performed on GRAPE6 boards \citep{mak02} located 
at the American Museum of Natural History and GPUs located at Swinburne University.

\subsection{The Star Clusters}

Each simulation started with $18\,000$ single stars and $2\,000$ binaries, 
i.e. $N = 20\,000$ objects comprising $22\,000$ stars and a 10\% primordial binary frequency. 
This is partly motivated by a desire to keep the models consistent with our 
earlier work in \citet{hur02}. 
In addition, $N \sim 2 \times 10^4$ is within the proposed range of cluster 
size (by star number and/or cluster mass) for the possible birth cluster of the Sun, albeit 
towards the upper end \citep{ada10}, and within the range suggested for the open 
cluster initial mass function \citep{pis08}. 
The binary frequency is on the low side for typical open clusters, with higher values 
to be explored in future simulations. 

A realistic initial-mass function is used to distribute the stellar masses \citep{kro93} 
between the limits of 0.1 and $50 \, M_\odot$, and we assume that the stars are of 
solar metallicity ($Z = 0.02$). The distribution of orbital separations for 
the primordial binaries is log-normal with a peak at $30\,$AU (see Eggleton, Fitchett \& Tout 1989), 
and spans the range $\sim 6 \, R_{\odot}$ to $30\,000\,$AU. 
The upper limit corresponds to an orbital period of $2 \times 10^9\,$d for a binary of 
solar mass, which matches the upper end of the period distribution for the nearby 
G-dwarf sample presented by Duquennoy \& Mayor (1991). 
Binary component masses are chosen according to a uniform distribution of mass-ratios.
The eccentricity of each binary orbit is taken from a thermal distribution (Heggie 1975). 
We use a Plummer density profile \citep{aar74} and assume the stars 
are in virial equilibrium when assigning the initial velocities.  
The cluster is subject to a standard Galactic tidal field 
-- a circular orbit at $8.5\,$kpc from the Galactic Centre -- 
with stars removed from the simulation when their distances from the density 
centre exceeds twice that of the tidal radius of the cluster. 
For our clusters with an initial mass of approximately $13\,200 \, M_\odot$ the initial 
tidal radius is $\sim 34\,$pc. 
We assume that all stars are on the zero-age main sequence (ZAMS) when 
the simulation begins and that any residual gas from the star formation 
process has been removed. 

We evolve a ``standard density" model which has an initial half-mass radius of $2.4\,$pc. 
In {\tt NBODY6} terms this means we set a length scale 
-- the free parameter in scaling from standard $N$-body units to physical units 
(Heggie \& Mathieu 1986) -- of  $R_{\rm sc} = 3\,$pc. 
The velocity dispersion of the starting model is $3.1 \,{\rm km } \,{\rm s}^{-1}$ and the 
initial half-mass relaxation timescale is about $140\,$Myr. 
The density of stars in the core of the cluster at the start of the 
simulation is $\sim 500\,$stars$\,{\rm pc}^{-3}$, with a steady decline to 
a fifth of this number after $\sim 200\,$Myr, and remaining approximately 
constant from that time onwards.  
The stellar density at the half-mass radius of the standard density model cluster is of order 
$50\,$stars$\,{\rm pc}^{-3}$ throughout. 

We also evolve a higher-density model which has an initial half-mass radius of $1.2\,$pc, 
i.e. $R_{\rm sc}$ is halved. 
The initial velocity dispersion and half-mass relaxation timescale are $4.4  \,{\rm km } \,{\rm s}^{-1}$ 
and $\sim 50\,$Myr, respectively. 
In this instance the density of stars in the core of the cluster at the start of the 
simulation is $\sim 4\,000\,$stars$\,{\rm pc}^{-3}$, dropping to a long-term value 
of  $\sim 1\,000\,$stars$\,{\rm pc}^{-3}$ from $200\,$Myr onwards. 

\subsection{The Planetary Systems}

In each simulation we include 100 planetary systems in the starting model. 
Each planetary system consists of two Jupiter-mass ($10^{-3} M_\odot$) planets.
We start with a Jupiter at its current distance from the Sun -- $5.2\,$AU -- (hereinafter the ``inner Jupiter"),
and add a second Jupiter located at the orbit of Saturn, at $9.5\,$AU from the Sun (the ``outer Jupiter").  
Initial eccentricities are 0.05 for each of the planets.
This corresponds to a separation between the Jupiters of about twice the dynamical stability limit (the ``Hills radius"). 
This relatively large separation ensures that any significant migrations will be due to the perturbations 
of passing stars rather than gravitational interactions between the planets.

The 100 parent stars containing planetary systems are chosen at random from the cluster stars except 
for the following restrictions. We require that the host stars' masses must be greater than 
$0.5 M_\odot$ and less than $1.1 M_\odot$. The upper limit ensures that the parent star stays on 
the main-sequence throughout the simulation. 
The lower limit of $0.5 M_\odot$ makes it less likely that the host stars and their
planetary systems quickly escape from the cluster. 
We also restrict the hosts to be single stars in the current study. 
Binary star hosts are possible and would likely increase the scope for orbital perturbations, 
given that binaries offer a larger cross-section for interactions with other stars than do 
single stars, but for simplicity (and efficiency of computation) we reserve this 
aspect for consideration in future work 
(see \citet{hao13} for a similar discussion). 

In essence we are evolving mini-solar-systems in an open cluster environment; 
owing to computational constraints we do not yet include more than two planets per star. 
Modelling full solar systems would require a dedicated solar system dynamics approach 
(e.g. Levison \& Duncan 1994; Thommes, Nagasawa \& Lin 2008). 
Instead, our planetary systems are treated as triple systems within the framework of 
{\tt NBODY6}, subject to stability conditions, regularization techniques, etc. as described 
in Aarseth (2003), with updates to the three-body stability algorithm given in Mardling (2008). 
To increase accuracy we do not go below Jupiter-mass for our planets for the time being 
and we use a reduced integration timestep for these low-mass objects. 
The fact that we have planetary systems that spend their lifetimes in the outskirts of 
our model clusters, essentially evolving in isolation, and that these systems do not 
exhibit variations of their orbital parameters points to the reliability of the method. 
Furthermore, we have independently verified the long-term stability of our chosen 
planetary systems by evolving them with the SWIFT orbit integrator (Levison \& Duncan 1994). 
Thus we are confident that any variations in planetary orbits reported in the following 
sections are the result of interactions within the star cluster environment rather than 
any intrinsic evolution of the planetary systems. 
However, we note that it would be prudent to employ even further reduced integration timesteps 
for planetary bodies in future simulations, particularly for perturbed orbits and even more so 
if bodies less massive than Jupiter are to be introduced, so as to err on the side of caution where small values of the binding energy are involved. 

We evolved five standard density simulations to $1\,$Gyr and continued two of these to $5\,$Gyr. 
We also evolved three high-density simulations to $1\,$Gyr. 
The age of $1\,$Gyr was chosen because it allowed sufficient time for the clusters to 
become dynamically old, with $\sim 10$ or more half-mass relaxation times elapsed, 
and because we expected any perturbation of the planetary systems to be more likely to occur 
early in the cluster evolution than later, noting the density decrease from the initial value. 
This latter assumption was checked (and borne out) by allowing two of the simulations to 
evolve to an age when clusters of this size are close to dissolution. 

\section{Results}

The $a$ versus $e$ plot for known exoplanets 
is a powerful constraint that must be explained by any successful planetary 
orbit evolution theory. The $a - e$ plots of the Jupiters in our standard and higher-density 
$N = 20\,000$ simulations are shown in Figures 1 and 2, respectively. 
A single point (circles for the inner Jupiters and crosses for the outer Jupiters) is placed on each figure
to represent the position in $a-e$ phase space for each planet every $0.2\,$Myr. 

In all standard simulations the outer Jupiters commonly migrate inwards to as close as 
$a = 7.5\,$ AU and outwards to as far as $a =12\,$AU.
The former value means that outer Jupiters reach close to the dynamical stability limit for inner Jupiters that have not migrated.
Some of the inner Jupiters migrate into the $3 - 6\,$AU range. 
It is therefore not unexpected that strong interactions 
leading to large planet migrations become likely in our simulations. 
In fact, outer Jupiters are subsequently found as close in as $3\,$AU. 
Furthermore, a key result of this paper is that three of the inner Jupiters reach $a = 0.33\,$AU 
($e = 0.985$), $a = 0.035\,$AU ($e = 0.616$), and $a = 0.012\,$AU 
($e = 0.2$) respectively, from their host stars, as detailed below. We emphasize that the formation of these three hot Jupiters is driven by passages of cluster stars near the planetary host stars, and that these events can {\it only} occur inside a star cluster. Solar systems born in the field with planets far from the dynamical instability, or ejected from clusters along with their host stars before stellar gravitational perturbations become significant, are impervious to this mechanism for hot Jupiter formation. 

As shown in Figure~2, the scattering within the higher-density simulations is even more 
pronounced, owing to the higher likelihood of close encounters between the stars hosting 
planets and other stars. The region between $3 - 15\,$AU is well populated by both inner and outer Jupiters. 
As a second key result we find an inner Jupiter at $a = 0.012\,$AU ($e = 0.202$) paired with an outer Jupiter 
at $a = 0.268\,$AU ($e = 0.582$) in the same planetary system. 
The system had evolved without interruption until just prior to the end of the 
second high-density simulation when, at $996\,$Myr, a close encounter with another star caused a strong perturbation which resulted 
in the observed parameters. The mass of the host star is $0.83 \, M_\odot$. 

Figure 3 follows the $a$ and $e$ histories of the Jupiters belonging to planetary system \#52 in our first standard simulation. 
The host star mass is $0.87 \, M_\odot$. The planetary system is unchanged until $t = 435\,$Myr, when the host star is located about 2 cluster core radii from the cluster center. The close passage of a  $0.60 \, M_\odot$ star perturbs both planets. This event, likely combined with an invoked short-period of chaotic evolution ending in a resonance, increases the eccentricity of the inner Jupiter to 0.25 while slightly decreasing its semi-major axis. The outer Jupiter's eccentricity is raised to 0.70 during this interaction, and its 
semi-major axis is increased nearly sixfold to $55\,$AU. For the next $370\,$Myr the eccentricity of the inner planet drifts downwards by a few percent while that of the outer planet rises to 0.74. 
For the entirety of this time interval the system resides either in or on the periphery of the 
cluster core, making it susceptible to a series of weak scattering events which also can 
contribute to manipulating the orbit. 
At around $800\,$Myr a $\sim 2 \, M_\odot$ star is nearby and a small increase in the centre-of-mass velocity of the planetary system is invoked. 
The orbit of the inner planet is then strongly perturbed by the outer planet 
and its motion becomes chaotic. 
Shortly afterwards at $t = 807\,$Myr the system actually escapes from the star cluster 
with the orbit of the inner planet having the remarkable values of $a = 0.035\,$AU and $e = 0.62$. 
The perihelion of the planet is a mere $2.8 \, R_\odot$, bringing it to within about two stellar 
radii of the photosphere of its host star every ``year". The outer planet has $a = 56\,$AU and $e = 0.74$ at the time of escape. 
We will note the similarities of the inner planet's orbit to those of HAT-P-2 b and HAT-P-34 b, amongst the closest exoplanets with nonzero eccentricity, in the next section. 

Figure 4 follows the $a$ and $e$ histories of the Jupiters belonging to planetary system \#79 in the second standard simulation. 
The host star mass is $0.59 \, M_\odot$. At $t = 58\,$Myr after the beginning of the simulation, the star is located about 3 cluster core radii from the cluster center. The close passage of a  $0.20 M_\odot$ star perturbs both planets, increasing the eccentricity of the inner Jupiter to 0.3 while barely changing its semi-major axis. The outer Jupiter's eccentricity is raised to 0.84 by the encounter, and its semi-major axis is increased nearly tenfold. For the next $130\,$Myr the eccentricities of these planets drift downwards by a few percent while their $a$ values are essentially constant. 
At $190\,$Myr a second but less intense stellar perturbation increases the outer planet's eccentricity to 0.9 while decreasing that of the inner Jupiter to 0.1. Just after $t = 400\,$Myr the planets begin to interact strongly, driving the eccentricity of the inner planet between 0.0 and 0.35 for about $80\,$Myr, while that of the outer planet rises to 0.985. During this chaotic period the semi-major axis of the outer planet reaches values as high as a few hundred AU or more and these are not shown in Figure~4 for clarity of scale. Finally the orbit stabilizes and at that 
point the inner planet 's semi-major axis has shrunk to $0.33\,$AU with a corresponding eccentricity of 0.985. 
The perihelion of the planet is a mere $1.4 \, R_\odot$, bringing it to within about one stellar radius of the photosphere of its host star at periastron. 
The outer planet stabilizes with $a = 58\,$AU and $e = 0.94$. 

Both planets remain fixed in these orbits, with an inclination that varies smoothly between $\sim 60$ and $\sim 130\,$degrees on 
a $\sim 10\,$Myr timescale, until we terminate the simulation at $1\,$Gyr. 
For the first $\sim 500\,$Myr of the evolution this system resides in or near the core of the cluster making it a target for weak scatters which contribute to changes in the orbit, similar to the previous example. However, for the remainder of the evolution the system has moved outwards to inhabit the less dense regions of the cluster. 
We will note the similarities of the inner planet's orbit to those of HD 80606 b -- the exoplanet with the largest known eccentricity -- in the next section.

The preceding two extreme exoplanets were formed over a period of about 500 Myr in our standard (low density cluster) simulations. 
One of our high density cluster simulations produced a third, remarkable hot Jupiter after 996 Myr: an inner Jupiter at $a = 0.012\,$AU ($e = 0.202$) paired with an outer Jupiter 
at $a = 0.268\,$AU ($e = 0.582$) in the same planetary system. This is the smallest semi major axis produced in any of our simulations, similar in size to smallest values of a ever detected for any exoplanet.

We can calculate the final separation of the inner planet in each of these three examples, after tidal circularization, 
by assuming conservation of angular momentum, $a(1-e^2)$ = constant (e.g. \citet{por01}). 
We find roughly 4.7, 2.1 and 2.5 R$_{\odot}$ for the three systems of note with (a,e) of (0.035,0.616), (0.33,0.985) and (0.012,0.202) respectively. 

Table~1 shows a census of the planetary systems in the standard and high-density 
clusters when the simulations end at $1\,$Gyr. Listed are the systems that remain in the cluster at that time and those that have escaped, 
averaged over the results for all simulations conducted. 
Also of interest is the longer-term outcome, hence the continuation of two of the standard 
simulations until an age of $5\,$Gyr when only 25\% (by mass) of the original cluster remained. 
The census for these simulations at $5\,$Gyr is also included in Table~1. 
We note that the $a-e$ plot for these simulations is not shown as it did not produce any notable 
systems beyond what has already been shown in Figure~1. 
 
As expected, the disruption of planetary systems is more efficient in the higher-density 
clusters and the incidence of perturbed systems is higher. In both cluster types the number of single-planet systems in existence at any point in 
time is small and similar. We can see from the $5\,$Gyr data for the standard clusters that given enough time the 
bulk of the planetary systems will escape from the cluster intact. 
Thus, perturbed systems of interest such as those highlighted in Figures~3 and 4 will be 
in the minority, even in higher density clusters. 
Hurley \& Shara (2002) found that $\sim 1\%$ of planets collided with their host star. 
The planetary systems in that study had a different morphology to those modelled here, 
being single-planet systems covering a wider range of orbits. 
However, we do find a similar incidence of collisions: 
about 0.5\% for the low-density simulations and 1\% for the high-density simulations. 

Also of interest is the presence of free-floating planets (FFP) both in 
the clusters and liberated to the host galaxy (see \citet{wan15} for recent N-body simulations which produce FFP).
Once again, as expected, the higher-density clusters produce more FFP. 
In a few cases the inner Jupiter is liberated but mostly it is the outer Jupiter. 
Interestingly, Hurley \& Shara (2002) found that about 10\% of their planetary systems 
produced a FFP and the numbers are comparable here even though the 
initial orbital configurations were quite different 
(and that our current  `systems' have two planets compared to their one). 
On average, one per simulation of these liberated planets even find themselves bound to a new host star. 

In their recent study of the survivability of multi-planet systems in open clusters via Monte Carlo scattering experiments, \citet{hao13} focussed on 4- and 5-planet systems evolving in an Orion Nebula Cluster-type environment for $\sim 100\,$Myr. Their planetary systems are not directly comparable to ours -- one system has Jupiter-mass planets at 1, 2.6, 6.5, 16.6 and $42.3\,$AU while the other has the four current gas giants of the Solar System. For the latter they show that the likelihood of the inner Jupiter surviving intact is 85\% which compares well with our 80-90\% (depending on the density of the cluster). \citet{hao13} do find that multiplicity decreases the survivability of a planetary system in a star cluster through induced planet-planet scattering, except in the case of the Jupiter in the mini-Solar System which was relatively unaffected by the presence of the outer lower-mass planets. This was verified by \citet{hao13} comparing against single-planet cases for each of the planets in their multi-planet systems. Using this approach they predict that a Saturn-mass planet orbiting at $9.5\,$AU has a 76\% chance of surviving when it is a single planet. This drops to 41\% when Jupiter, Neptune and Uranus are included. Our planet at $9.5\,$AU has a $\sim 90$\% chance of surviving, but of course has Jupiter-mass and does not have planets exterior to it, with both factors increasing the chance of survival.

\section{Comparison with Known Extreme Exoplanets}

The exoplanet with the largest well-determined eccentricity is HD 80606b; $e = 0.9332$ and the remarkably small $a = 0.449\,$AU \citep{nai01,pon09}. These same authors offered several suggestions for the origin of HD 80606b's orbit, including the gravitational interaction with another planet(s), especially if that other planet were expelled in a violent interaction. The orbit of the simulated inner planet of Figure 3 is a remarkable match to HD 80606b, with $a = 0.33\,$AU and $e = 0.985$. If the formation scenario of HD 80606b is at all similar to that detailed in the previous section and in Figure 3 then there may well be a second planet in the HD 80606 system with high eccentricity ($e \geq 0.9$ ) and $a \approx 60\,$AU. 

Significantly eccentric orbits at even the smallest semi-major axes are seen in $\approx$ 30\% of the known exoplanets. At a semi-major axis of $0.0674\,$AU we encounter an eccentricity of 0.5171 for HAT-P-2b \citep{bak07}. The simulation depicted in Figure 4 produces an exoplanet with a similar orbit; with $a = 0.034\,$AU and $e = 0.62$, it is slightly more ``extreme" than HAT-P-2b. Just as in the previous example, we would expect to find a second massive planet with $a \approx 70\,$AU and $e \geq 0.7$ in the HAT-P-2 system if the formation mechanism is similar to that detailed in Figure 4. \citet{lew13} have recently cited observational evidence for the companion we predict, noting evidence for a long-term linear trend in the star's radial velocity data.

As noted above, one of our high density cluster simulations produced an inner Jupiter at $a = 0.012\,$AU ($e = 0.202$) paired with an outer Jupiter 
at $a = 0.268\,$AU ($e = 0.582$) in the same planetary system. We are unaware of a close analog (yet!) amongst the known exoplanets to such a system, but we note the orbital similarities presented by the brown dwarf HD 41004 B b ($a = 0.0177\,$AU and $e = 0.081$) \citep{zuc04}, and by the super-Earth GJ 876 d ($a = 0.0208\,$AU and $e = 0.207$) \citep{riv10}.

\section{How Common are Hot Jupiters?}
We produced three extremely hot Jupiters in 8 simulations of star clusters of 22,000 stars with 10\% binary fractions -- a total of 160,000 stellar systems. For ease of computation, only 100 of the 20,000 systems in each simulation was populated with two Jupiters. The fraction of low and moderate mass stars with planets is of order 100\% \citep{cla14}. Thus if our simulations had started with 20,000 pairs of Jupiters in each cluster instead of just 100 pairs we would expect to have produced of order 3 x 200 = 600 hot Jupiters. 
This corresponds to 600/160,000 = 0.4\% of all stellar systems 
(noting that we are extrapolating over all masses but only $\sim 10\%$ of the stars in our models have masses greater than $1.1 \, M_\odot$). 
Certainly in the cases of NGC 188 \citep{gel12} and M67 \citep{lat07} the observed binary fraction is much closer to 30\%, increasing our (very rough) guesstimate to 1.2\% of all stellar systems. The larger cross-sections of binaries, and a higher binary fraction will inevitably lead to more perturbations of planetary systems, and yet more hot Jupiters. Finally, the presence of three or more planets must also increase the fraction of hot Jupiters. Gravitational perturbations by passing stars thus seem capable of yielding hot Jupiters at a frequency of order 1\% of all planetary systems in populous open star clusters with binary fractions similar to those of M67 and NGC 188. There is growing evidence that, just as our simulations predict, distant companions accompany hot Jupiters \citet{knu14}.

Finally we emphasize that the extreme hot Jupiters that our simulations produce are not meant to be taken as detailed models of any particular exoplanet. Rather, the ease with which we can produce very hot Jupiters in highly eccentric orbits in star clusters is the key result of this paper.
Many more simulations (varying the masses, initial separations and numbers of exoplanets in star clusters with different initial conditions) will be essential to making more concrete, testable predictions about the distributions of expected orbital parameters of exoplanets produced by dynamical interactions in clusters. 

For these future simulations we should also employ a finer time resolution for the output snapshots 
of the cluster members and additional documentation of close encounters. 
Such measures increase the overheads of course, in terms of cpu-time and the size of data files, 
but are necessary if we wish to pinpoint the nature of the interactions leading to orbital changes 
in greater detail. 
Taking even greater care with the integration of perturbed low-mass bodies, given the small energies 
involved and wanting to ensure that any errors are well below these in magnitude, will also 
introduce cpu-time overheads but will help to further validate the process. 
 
\section{Conclusions}
\label{s:conclu}

We have used direct $N$-body integrations to observe the fates of 100 planetary systems 
(composed of two Jupiters) in moderately dense open clusters with $N = 18\,000$ single stars and $2\,000$ binaries. 
Interactions with passing stars lead, on less than a Gyr timescale, to strong interactions between the two planets. 
The subsequent dynamical evolution produces three very hot Jupiters with orbits that mimic those of the
exoplanets with the most extreme semi-major axes and eccentricities known. 
Our simulations of the stellar perturbations that formed these hot Jupiters predict that very hot, inner planets are 
likely to be accompanied by much more distant gas giants in highly eccentric orbits.
Being born in a star cluster can certainly mold planetary systems into even the most extreme exoplanet-like systems. 
Significantly more exploration of planetary system and star cluster conditions will be needed to determine if a majority
of observed exoplanetary systems' orbits have been modified by stellar encounters 
and to further verify this mechanism for producing hot Jupiters in extreme configurations. 

\acknowledgments
 
 MMS gratefully acknowledges the support of Hilary and Ethel Lipsitz, good friends and supporters of the 
 Department of Astrophysics of the American Museum of Natural History. 
 Part of this work was performed on the gSTAR national facility at Swinburne University of Technology. 
 gSTAR is funded by Swinburne and the Australian GovernmentÕs Education Investment Fund.

\clearpage

\begin{figure}
\plotone{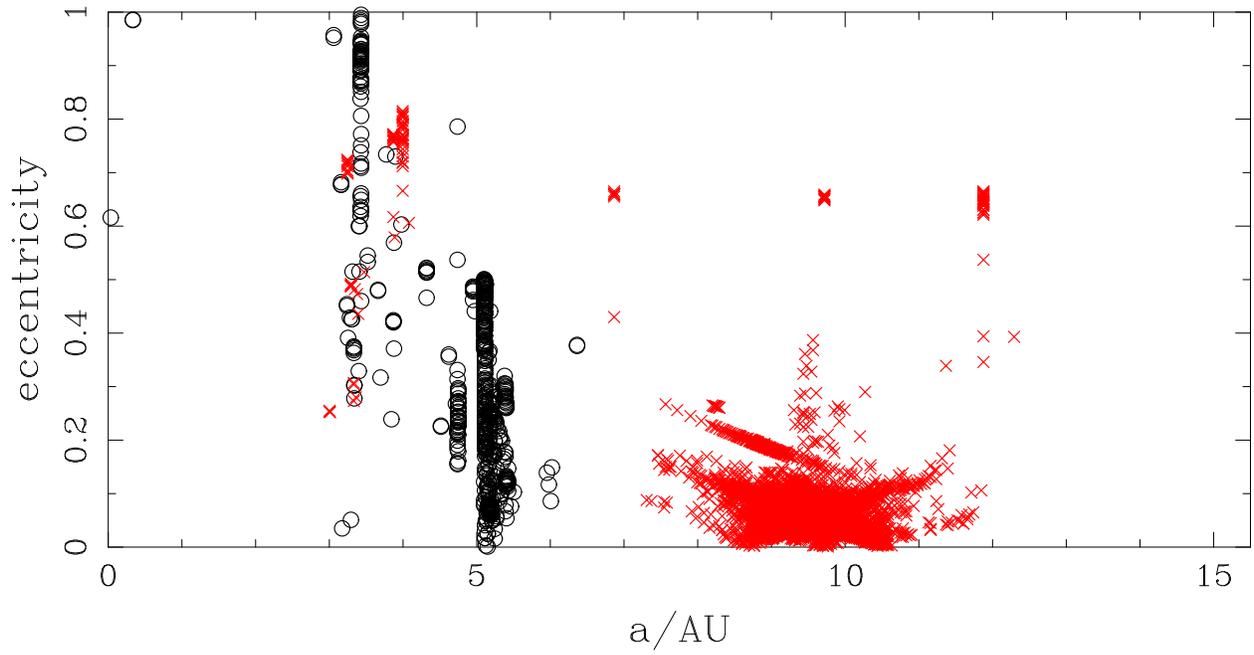}
\caption{
Eccentricity versus semi-major axis for all bound Jupiters in the standard set of $N = 20\,000$ open clusters (five simulations in total). 
The outer Jupiter-mass planet (starting at $9.5\,$AU) is shown as red crosses and the 
inner Jupiter (starting at $5.2\,$AU) as black open circles.
Orbital parameters are shown every $0.2\,$Myr up to a cluster age of $1\,$Gyr. 
\label{f:fig1}}
\end{figure}

\clearpage

\begin{figure}
\plotone{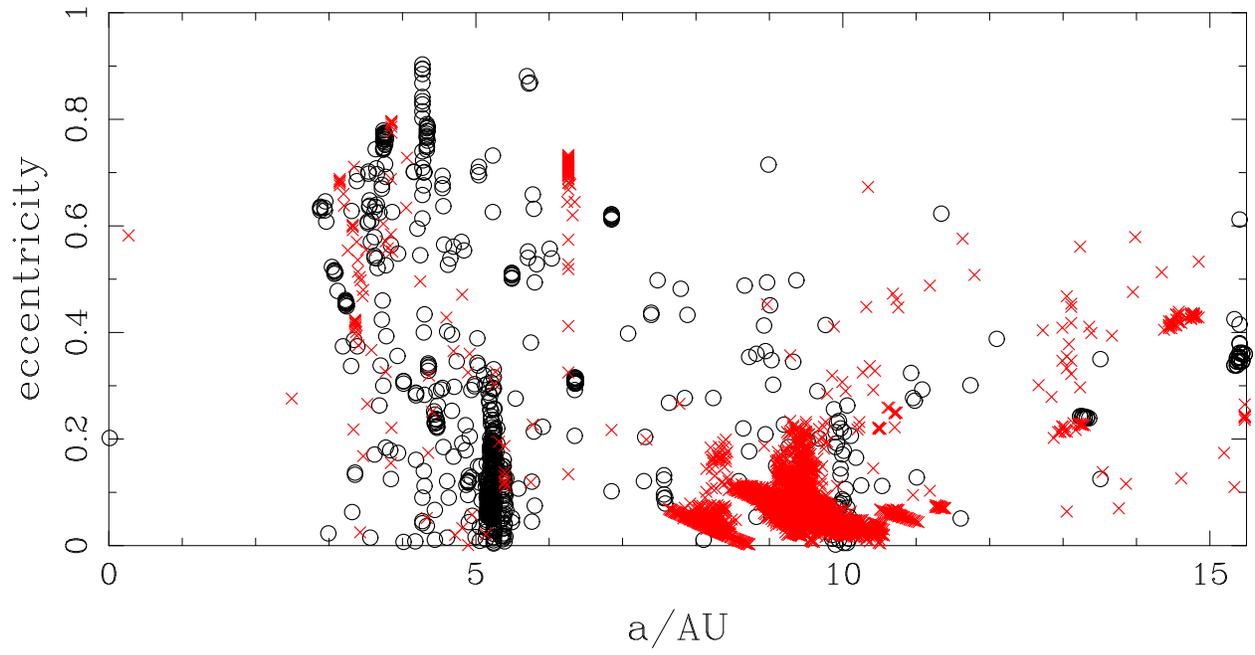}
\caption{Same as Figure~\ref{f:fig1} but for the higher density set of cluster models (three simulations in total). 
 \label{f:fig2}}
\end{figure}

\clearpage

\begin{figure}
\plotone{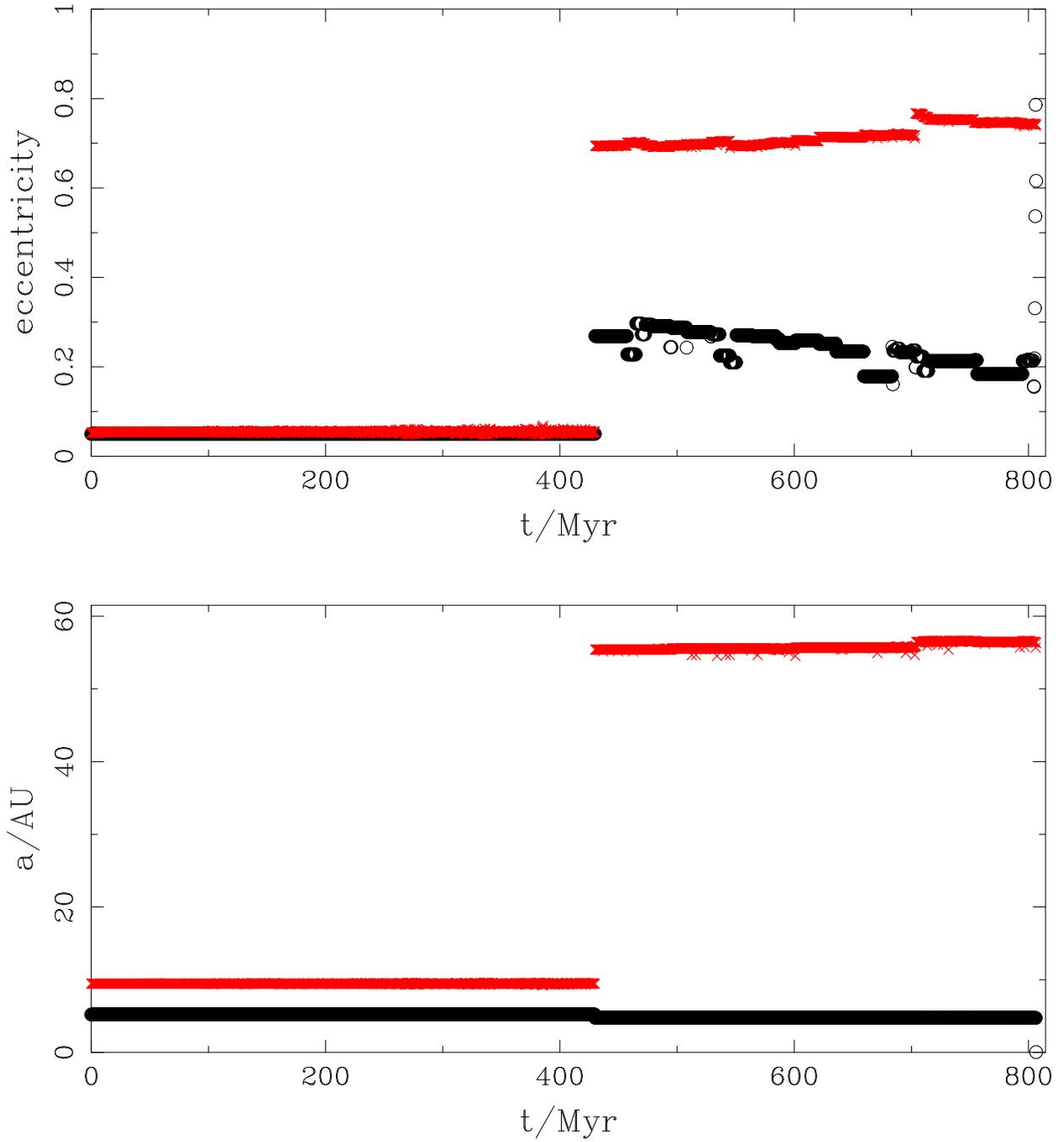}
\caption{
Eccentricity versus time (upper panel) and semi-major axis versus time (lower panel) for  
planetary system \#52 of the first standard simulation.
Orbital parameters are shown every $0.2\,$Myr. 
The host star is $0.87 M_\odot$. 
 \label{f:fig3}}
\end{figure}

\clearpage

\begin{figure}
\plotone{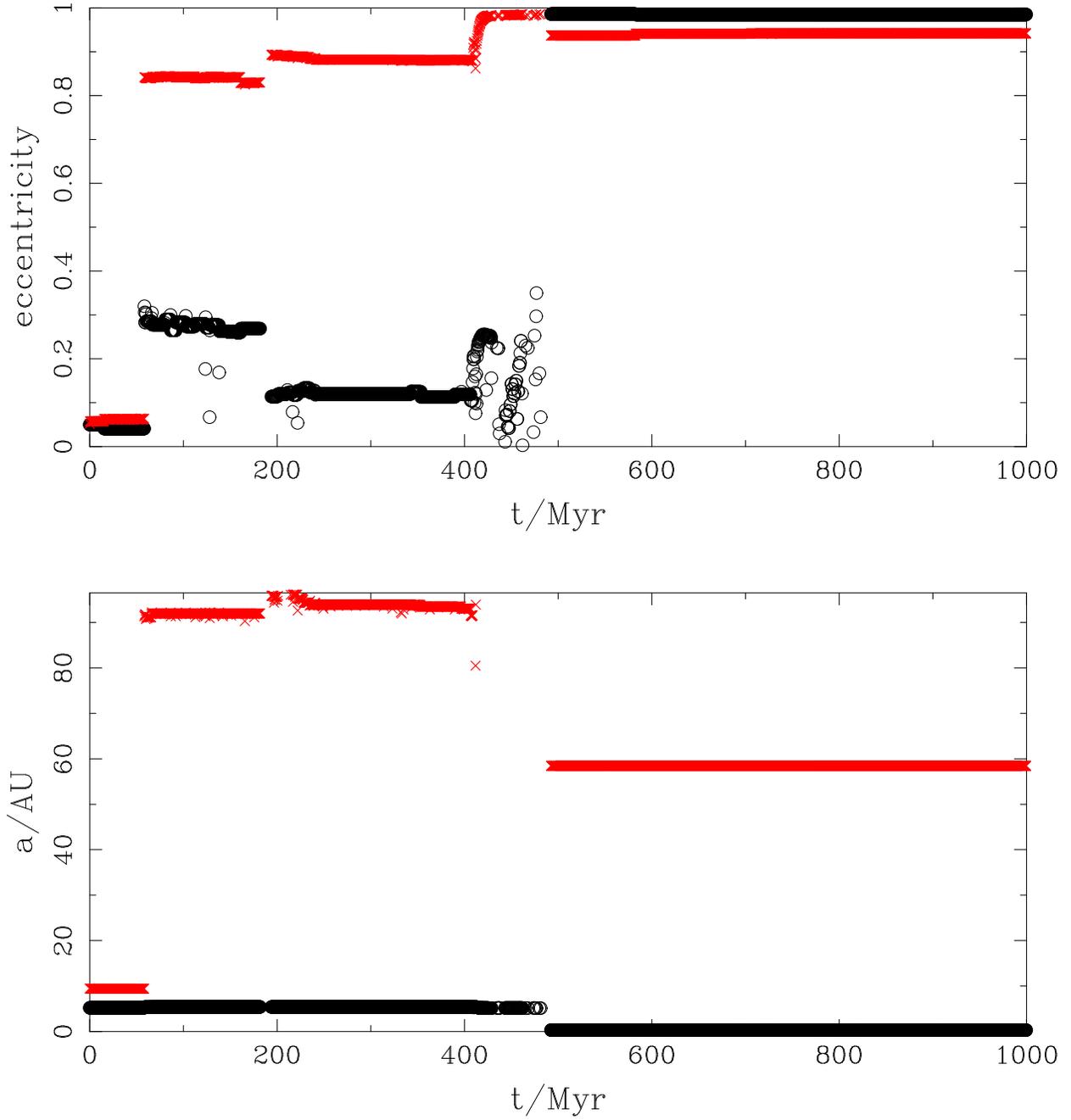}
\caption{
Same as Figure~\ref{f:fig3} but for planetary system \#79 of the second standard simulation. 
The host star is $0.59 M_\odot$. 
\label{f:fig4}}
\end{figure}

\clearpage

\begin{deluxetable}{lrrr}
\tablecolumns{4}
\tablewidth{0pc}
\tablecaption{
Census of planetary systems remaining in a cluster or having escaped by a particular time. 
The first column provides labels of the systems included in the census: $*12$ denotes a planetary system (host star and two Jupiters) that has remained intact and unperturbed; 
$*12'$ denotes a planetary system that remains intact but has been perturbed (at least one orbital parameter has changed by more then 10\%); 
$*1$ is a host star and inner Jupiter only; 
$*2$ is a host star and outer Jupiter only; 
$*$ denotes a host star that has lost both planets; 
and FFP refers to planets unbound from a host star. 
Systems remaining in the cluster at the specified time are listed first, followed by systems that 
have escaped in the lower rows. 
Column 2 shows data for the standard simulations after $1\,$Gyr (average of five simulations), 
column 3 shows data for the higher-density simulations after $1\,$Gyr (average of three simulations) and the final column is for the standard simulations after $5\,$Gyr 
(average of two simulations). 
\label{t:table1}
}
\tabletypesize\normalsize
\tablehead{
%  & S1 & H1 & S5 
  & standard & high-d & standard \\
  & $1\,$Gyr & $1\,$Gyr & $5\,$Gyr 
}
\startdata 
& \multicolumn{3}{c}{remaining in the cluster}\\
$*12$ & 88 & 76 & 38 \\ 
$*12'$ & 1 & 7 & 2 \\
$*1$ & 3 & 3 & 3 \\
$*2$ & 1 & 1 & 2 \\
$*$ & 1 & 3 & 1 \\
FFP & 4 & 5 & 2 \\ 
 &  \multicolumn{3}{c}{escaped from the cluster}  \\
$*12$ & 5 & 7 & 50 \\
$*12'$ & 1 & 1 & 2 \\
$*1$ & 0 & 1 & 1 \\
$*2$ & 0 & 1 & 1 \\
FFP & 2 & 7 & 7 \\
\enddata
\end{deluxetable}


\newpage 

\begin{thebibliography}{}
\bibitem[Aarseth, H\'{e}non \& Wielen(1974)]{aar74} Aarseth, S.,
    H\'{e}non, M., \& Wielen, R.1974, \aap, 37, 183 
\bibitem[Aarseth (1999)]{aar99} Aarseth, S. J. 1999, \pasp, 111, 1333 
\bibitem[Adams \& Laughlin (2001)]{ada01} Adams, F.C., \& Laughlin, G. 2001, 
    Icarus, 150, 151 
\bibitem[Adams(2005)]{ada05} Adams, F.~C.\ 2005, IAU Colloq.~197: Dynamics of Populations of Planetary Systems, 19 
\bibitem[Adams \& Laughlin(2006)]{ada06} Adams, F.~C., \& Laughlin, G.\ 2006, \apj, 649, 1004 
\bibitem[Adams (2010)]{ada10} Adams, F.~C. 2010, \araa, 48, 47
\bibitem[Bakos et al.(2007)]{bak07} Bakos, G.~{\'A}., 
Kov{\'a}cs, G., Torres, G., et al.\ 2007, \apj, 670, 826 
\bibitem[Batalha et al.(2013)]{bat13} Batalha, N.~M., Rowe, 
J.~F., Bryson, S.~T., et al.\ 2013, \apjs, 204, 24 
\bibitem[Boley et al.(2012)]{bol12} Boley, A.~C., Payne, M.~J., \& Ford, E.~B.\ 2012, \apj, 754, 57 
\bibitem[Bonnell et al. (2001)]{bon01} Bonnell, I. A., Smith, K. W., Davies, M. B., \& Horne, K. 2001, \mnras, 322, 859
\bibitem[Brucalassi et al.(2014)]{bru14} Brucalassi, A., Pasquini, L., Saglia, R., et al.\ 2014, \aap, 561, L9 
\bibitem[Clanton \& Gaudi(2014)]{cla14} Clanton, C., \& Gaudi, B.~S.\ 2014, \apj, 791, 91 
\bibitem[Cochran, Hatzes \& Paulson (2002)]{coc02} Cochran, W.D., 
    Hatzes, A.P., \& Paulson, D.B. 2002, \aj, 124, 565
\bibitem[Davies \& Sigurdsson (2001)]{dav01} Davies, M. B., 
    \& Sigurdsson, S. 2001, \mnras, 324, 612
\bibitem[de Juan Ovelar et al.(2012)]{dej12} de Juan Ovelar, M., Kruijssen, J.~M.~D., Bressert, E., et al.\ 2012, \aap, 546, LL1 
\bibitem[Duquennoy \& Mayor (1991)]{duq91} Duquennoy, A., \& Mayor, M.\ 1991, \aap, 248, 485
\bibitem[Eggenberger, Udry \& Mayor (2003)]{egg03} Eggenberger, A., Udry, S., \& Mayor, M. 2003, 
 in ASP Conference Series 294, Scientific Frontiers in Research on Extrasolar Planets, 
 ed. D. Deming \& S. Seager (San Francisco: ASP), 43
\bibitem[Eggleton, Fitchett \& Tout (1989)]{egg89} Eggleton, P.P., Fitchett, M., 
    \& Tout, C.A. 1989, \apj, 347, 998 
\bibitem[Ford, Rasio \& Yu (2003)]{for03} Ford, E.B., Rasio, F.A., \& Yu, K. 2003, in ASP Conference Series 294, 
Scientific Frontiers in Research on Extrasolar Planets, ed. D. Deming \& S. Seager (San Francisco: ASP), 181
\bibitem[Ford et al.(2006)]{for06} Ford, E.~B., Rasio, F.~A., 
\& Yu, K.\ 2006, ISSI Scientific Reports Series, 6, 123 
\bibitem[de la Fuente Marcos \& de la Fuente Marcos (1997)]{fue97} de la Fuente Marcos, C.,
    \& de la Fuente Marcos R. 1997, \aap, 326, L21
\bibitem[Geller \& Mathieu(2012)]{gel12} Geller, A.~M., \& Mathieu, R.~D.\ 2012, \aj, 144, 54 
\bibitem[Goddard et al.(2010)]{god10} Goddard, Q.~E., Bastian, N., \& Kennicutt, R.~C.\ 2010, \mnras, 405, 857 
\bibitem[Goldreich \& Tremaine (1980)]{gol80} Goldreich, P., \& Tremaine, S. 1980, \apj, 241, 425
\bibitem[Gratton et al. (2001)]{gra01} Gratton, R.G., Bonanno, G., Claudi, R.U., 
    Cosentino, R., Desidera, S., Lucatello, S., \& Scuderi, S. 2001, \aap, 377, 123 
\bibitem[Hao et al.(2013)]{hao13} Hao, W., Kouwenhoven, M.~B.~N., \& Spurzem, R.\ 2013, \mnras, 433, 867 
\bibitem[Heggie \& Rasio (1996)]{heg96} Heggie, D.C., \& Rasio, F.A. 1996, \mnras, 282, 1064
\bibitem[Hurley \& Shara (2002)]{hur02} Hurley, J.R., \& Shara, M.M. 2002, \apj, 565, 1251
\bibitem[Hurley et al. (2001)]{hur01} Hurley, J. R., Tout, C. A., Aarseth, S. J., \& Pols, O.R. 2001, \mnras, 323, 630
\bibitem[Hurley et al.(2008)]{hsm08} Hurley, J.~R., Shara, M.~M., \& Mardling, R.~A.\ 2008, Extreme Solar Systems, 398, 137 
\bibitem[Israelian et al. (2001)]{isr01} Israelian, G., Santos, N.C., Mayor, M., \& Rebolo, R. 2001, \nat, 411, 163 
\bibitem[Knutson et al.(2014)]{knu14} Knutson, H.~A., Fulton, B.~J., Montet, B.~T., et al.\ 2014, \apj, 785, 126 
\bibitem[Kroupa, Aarseth \& Hurley (2000)]{kro00} Kroupa, P., Aarseth, S., \& Hurley, J. 2000, \mnras, 321, 699
\bibitem[Kroupa, Tout \& Gilmore (1993)]{kro93} Kroupa, P., Tout, C. A., \& Gilmore, G. 1993, \mnras, 262, 545
\bibitem[Kruijssen(2012)]{kru12} Kruijssen, J.~M.~D.\ 2012, \mnras, 426, 3008 
\bibitem[Lada \& Lada(2003)]{lad03} Lada, C.~J., \& Lada, E.~A.\ 2003, \araa, 41, 57 
\bibitem[Latham(2007)]{lat07} Latham, D.~W.\ 2007, Highlights of Astronomy, 14, 444 
\bibitem[Laughlin \& Adams (1998)]{lau98} Laughlin, G., \& Adams, F.C. 1998, \apj, 508, L171
\bibitem[Lewis et al.(2013)]{lew13} Lewis, N.~K., Knutson, H.~A., Showman, A.~P., et al.\ 2013, \apj, 766, 95 
\bibitem[Li \& Adams(2015)]{ada15} Li, G., \& Adams, F.~C.\ 2015, \mnras, 448, 344 
\bibitem[Lin, Bodenheimer  \& Richardson (1996)]{lin96} Lin, D.N.C., Bodenheimer, P., \& Richardson, D.C. 1996, \nat, 380, 606 
\bibitem[Lin \& Dobbs-Dixon(2008)]{lin08} Lin, D.~N.~C., \& Dobbs-Dixon, I.\ 2008, IAU Symposium, 249, 131    
\bibitem[Lovis \& Mayor(2007)]{lov07} Lovis, C., \& Mayor, M.\ 2007, \aap, 472, 657 
\bibitem[Makino (2002)]{mak02} Makino, J. 2002, in ASP Conference Series 263, 
 Stellar Collisions, Mergers and their Consequences, ed. M.M. Shara  (San Francisco: ASP), 389 
\bibitem[Malmberg et al.(2011)]{mal11} Malmberg, D., Davies, M.~B., \& Heggie, D.~C.\ 2011, \mnras, 411, 859 
\bibitem[Marcy et al. (2003)]{mar03} Marcy, G.W., Butler, R.P., Fischer, D.A., \& Vogt, S.S. 2003, in ASP Conference Series 294, 
Scientific Frontiers in Research on Extrasolar Planets, ed. D. Deming \& S. Seager (San Francisco: ASP), 1
\bibitem[Mardling (2008)]{mar08} Mardling, R.A. 2008, in Proceedings of IAU Symposium 246, 
Dynamical Evolution of Dense Stellar Systems, ed. E. Vesperini, M. Giersz \& A. Sills  (Cambridge University Press), 199
\bibitem[Mayor \& Queloz(1995)]{may95} Mayor, M., \& Queloz, D.\ 1995, \nat, 378, 355 
\bibitem[Meibom et al.(2013)]{mei13} Meibom, S., Torres, G., Fressin, F., et al.\ 2013, \nat, 499, 55 
\bibitem[Murray (2003)]{mur03} Murray, N. 2003, in ASP Conference Series 294, Scientific Frontiers in Research on Extrasolar Planets, 
ed. D. Deming \& S. Seager (San Francisco: ASP), 165 
\bibitem[Murray and Dermott(1999)]{mur99} Murray, C.D. \& Dermott, S.F.\ 1999,  Solar System Dynamics, Cambridge University Press, Cambridge
 \bibitem[Naef et al.(2001)]{nai01} Naef, D., Latham, D.~W., Mayor, M., et al.\ 2001, \aap, 375, L27 
 \bibitem[Parker \& Quanz(2012)]{par12} Parker, R.~J., \& Quanz, S.~P.\ 2012, \mnras, 419, 2448  
 \bibitem[Piskunov et al.(2008)]{pis08} Piskunov, A.E., Kharchenko, N.V., Schilbach, E., R\"{o}ser, S., Scholz, R.-D., \& Zinnecker, H.\ 2008, \aap, 487, 557
\bibitem[Pollack et al. (1996)]{pol96} Pollack, J.B., Hubickyj, O., Bodenheimer, P., Lissauer, J.L., Podolak, M., \& Greenzweig, Y. 1996, Icarus, 124, 62
\bibitem[Pont et al.(2009)]{pon09} Pont, F., H{\'e}brard, G., Irwin, J.~M., et al.\ 2009, \aap, 502, 695 
\bibitem[Portegies Zwart et al.(2001)]{por01} Portegies Zwart, S.~F., McMillan, S.~L.~W., Hut, P., \& Makino, J.\ 2001, \mnras, 321, 199 
\bibitem[Pr{\v s}a et al.(2011)]{prs11} Pr{\v s}a, A., Batalha, N., Slawson, R.~W., et al.\ 2011, \aj, 141, 83 
\bibitem[Quinn et al.(2012)]{qui12} Quinn, S.~N., White, R.~J., Latham, D.~W., et al.\ 2012, \apjl, 756, LL33 
\bibitem[Quinn et al.(2014)]{qui14} Quinn, S.~N., White, R.~J., Latham, D.~W., et al.\ 2014, \apj, 787, 27 
\bibitem[Rasio \& Ford (1996)]{ras96} Rasio, F.A., \& Ford, E.B. 1996, Science, 274, 954
\bibitem[Rivera et al.(2010)]{riv10} Rivera, E.~J., Laughlin, G., Butler, R.~P., et al.\ 2010, \apj, 719, 890 
\bibitem[Sato et al.(2007)]{sat07} Sato, B., Izumiura, H., Toyota, E., et al.\ 2007, \apj, 661, 527
\bibitem[Spurzem \& Lin (2003)]{spu03} Spurzem, R., \& Lin, D.N.C. 2003, in ASP Conference Series 294, 
Scientific Frontiers in Research on Extrasolar Planets, ed. D. Deming \& S. Seager (San Francisco: ASP), 217
\bibitem[Spurzem et al.(2009)]{spu09} Spurzem, R., Giersz, M., Heggie, D.~C., \& Lin, D.~N.~C.\ 2009, \apj, 697, 458 
\bibitem[Thies et al.(2011)]{thi11} Thies, I., Kroupa, P., Goodwin, S.~P., Stamatellos, D., \& Whitworth, A.~P.\ 2011, \mnras, 417, 1817 
\bibitem[Thorsett et al. (1999)]{tho99} Thorsett, S. E., Arzoumanian, Z., Camilo, F., \& Lyne, A. G. 1999, \apj, 532, 763
\bibitem[Wang et al.(2015)]{wan15} Wang, L., Kouwenhoven, M.~B.~N., Zheng, X., Church, R.~P., \& Davies, M.~B.\ 2015, \mnras, 449, 3543 
\bibitem[Wolszczan \& Frail (1992)]{wol92} Wolszczan, A., \& Frail, D.D. 1992, \nat, 355, 145
\bibitem[Zucker et al.(2004)]{zuc04} Zucker, S., Mazeh, T., Santos, N.~C., Udry, S., \& Mayor, M.\ 2004, \aap, 426, 695 
\end{thebibliography}
\end{document}